\documentclass{article}
\usepackage{amsmath}
\usepackage{amssymb}
\usepackage{graphicx}
\begin{document}

\begin{center}
\LARGE
\textbf{Quantum key distribution}

\textbf{without the wavefunction}
\\[0,5 cm]
\normalsize
Gerd Niestegge
\\[0,5 cm]
\footnotesize
Zillertalstrasse 39, 81373 M\"unchen, Germany

gerd.niestegge@web.de
\\[0,5 cm]
\end{center}
\normalsize
\begin{abstract}
A well-known feature of quantum mechanics
is the secure exchange of secret bit strings
which can then be used as keys to encrypt messages 
transmitted over any classical
communication channel. 
It is demonstrated that this
quantum key distribution allows a much more general and abstract access than
commonly thought. The results include some
generalizations for the Hilbert space version of quantum key distribution,
but base upon a general
non-classical extension of conditional probability.
A special state-independent conditional probability is 
identified as origin 
of the superior security of quantum key distribution
and may have more profound implications for the foundations
and interpretation of quantum mechanics, 
quantum information theory, and the philosophical question
what actually constitutes physical reality.
\\[0,2 cm]
\\[0,2 cm]
\\[0,2 cm]
\end{abstract}
\noindent
\large
\textbf{1. Introduction}
\normalsize
\\[0,5 cm]
In the past thirty years, quantum cryptography has been a large field of
extensive theoretical and experimental research.  
Several quantum key distribution protocols have been elaborated,
their technical feasibility has been discussed,  
experimental set-ups and some practical implementations
have been realized. Quantum key distribution may be the first 
commercial application of quantum information theory.
A vast amount of papers on these topics are available by now
(Each of the two pioneering works \cite{BB84} and \cite{E91} has more
than ten thousand and almost eight thousand quotations, respectively; 
for a review, see \cite{gisin2002cryptoreview} 
or \cite{2009cryptoreview}).
The present paper, however, 
focuses on the theoretical foundations and 
concentrates on the two best-known 
quantum key distribution protocols.

Quantum mechanics renders possible the secure exchange of secret bit strings 
which can then be used as keys to encrypt messages transmitted over a 
classical communication channel. The quantum no-cloning theorem 
\cite{dieks1982communication, wootters1982single, yuen1986amplification} ensures
that an eavesdropper cannot steal the key without being detected. 
These results follow from the usual quantum mechanical 
Hilbert space formalism.
After it could recently be shown in Ref.\cite{Nie2015PhysScrNoCloning} 
that the quantum no-cloning theorem 
can be extended to a much more general and abstract framework - the quantum 
logics which possess a conditional probability calculus -, a similar
extension of quantum key distribution shall be presented in this paper.

A quantum logic $E$ is a purely algebraic structure. 
It is quite common to use an orthomodular 
partially ordered set or lattice \cite{beltrametti1984logic, 
beran1985orthomodular, kalmbachorthomodular, ptak1991orthomodular}.
Its elements can be interpreted in different ways: as events, 
propositions or properties of the 
quantum system under consideration.  
This is why they are called propositions by some authors
and events, preferably in probabilistic approaches, by others.

States are then defined in the same way as the classical 
probability measures, and conditional probabilities are 
postulated to behave like the classical ones on compatible 
subsets of $E$. Note that a subset is called compatible 
if it is contained in another subset of $E$ forming 
a Boolean algebra (i.e., it is contained in a classical subsystem of $E$) 
\cite{brabec1979compatibility}. 

Some quantum logics 
entail unique conditional probabilities, many others don't. 
The classical Boolean algebras and the Hilbert space quantum logic 
(consisting of the closed subspaces or, equivalently, 
the selfadjoint projection operators) do and, 
in the latter case, conditionalization becomes identical 
with the state transition of the 
L\"uders - von Neumann (i.e., projective) quantum 
measurement process \cite{niestegge2001non}. Therefore, 
the quantum logics which possess unique conditional probabilities 
can be regarded as a generalized model 
of projective quantum measurement.
Examples that are neither classical Boolean algebras nor 
Hilbert space quantum logics (nor sublogics of them) can be 
constructed using the exceptional real Jordan algebras 
\cite{niestegge2001non, niestegge2008approach}. 

In this framework, a very special type of conditional probability emerges 
in the non-classical case \cite{niestegge2001non, niestegge2008approach}. 
It describes the probability for the transition from a past event $e$ to a future event $f$,
independently of any underlying state, and results solely from the algebraic structure 
of the quantum logic $E$. This probability exists only for certain pairs $e$ and $f$ in $E$. 
It exists for all $f \in E$, if $e$ is minimal in $E$ (i.e., $e$ is an \textit{atom}). 
The states resulting in this way are called \textit{atomic}. They represent a generalization 
of the pure states in the usual quantum mechanical Hilbert space setting.

After the early pioneering work by Birkhoff and von Neumann in 1936 \cite{birkhoff-vN36}, 
quantum logics have been studied extensively between 1960 and 1995 
\cite{beltrametti1984logic, beran1985orthomodular, kalmbachorthomodular, Keller1980, 
piron1964axiomatique, ptak1991orthomodular, soler1995characterization, 
varadarajan1968geometry, varadarajan1970geometry}. 
Various forms of conditional probability have also been considered
\cite{beltrametti1984logic, bub1977neumann, edwards1990conditional, 
friedman1978quantum, gunson1967algebraic, guz1981conditional}.  
However, the quantum logics which possess unique conditional probabilities and particularly
the special type of the state-independent conditional probability 
had not received any attention before the author's work 
\cite{Nie1998HPA01, niestegge2001non, niestegge2008approach}.

This special type of conditional probability 
is used in the present paper to study
two well-known quantum key distribution protocols
(including a variant of one of them). The first one was
invented by Charles H. Bennett and Gilles Brassard in 1984 \cite{BB84}, based
on earlier work in the 1970ies by Stephen Wiesner 
which was not published before 1983 \cite{wiesner1983}.
Its variant is due to the first co-inventor \cite{bennett1992nonortho}.
The second protocol was proposed by Artur K. Ekert in 1991 \cite{E91}.

These quantum key distribution protocols are generalized in a two-fold way.
First, they are transferred to 
the quantum logics possessing the conditional probability calculus; this 
shows that quantum key distribution does not necessarily require the full-blown Hilbert 
space formalism of quantum mechanics and works as well in a much more 
general and abstract model of the projective quantum measurement process. 
Second, system properties 
are used instead of states. Pure states, wavefunctions 
and their collapse are not needed.

Before turning to quantum key distribution, the no-cloning theorem 
of Ref. \cite{Nie2015PhysScrNoCloning} is revisited and extended
by treating the cloning of system properties, since only the 
cloning of atomic states was considered in Ref. \cite{Nie2015PhysScrNoCloning}.

Moreover, the special type of conditional probability mentioned above 
is identified as the quantum origin of the no-cloning theorem as well as 
of the superior security of the key distribution processes.
Other generalizations of 
the no-cloning theorem \cite{barnum2006cloning, barnum2007generalized}
and quantum key distribution \cite{barrett2007information},
using the generalized probabilistic theories,
do not cover the results presented here,
since they do not capture this special type of conditional probability.

The paper is organized as follows. The next two sections 
restate some material from Refs.
\cite{niestegge2001non, niestegge2008approach, Nie2015PhysScrNoCloning} 
as far as needed in the present paper. In section 2,
the algebraic structure of a quantum logic is outlined and, in section 3, the 
conditional probability calculus. In section 4, the extension of the no-cloning theorem is
introduced. The generalized versions of the
quantum key distribution protocols are presented in section 5.
\\[0,5 cm]
\large
\textbf{2. Compatibility in orthomodular partially ordered sets}
\\[0,5 cm]
\normalsize
In quantum mechanics, the measurable quantities of a physical 
system are re\-presented by observables. Most simple are those 
observables where only the two values `true' and `false'
(or `1' and `0')
are possible as measurement outcome.
They are elements of a mathematical structure called \textit{quantum logic},
are usually called \textit{propositions}, 
and they are called \textit{events} in probabilistic approaches.
The elements of the quantum logic can also be understood as potential properties 
of the system under consideration. If, for instance, this system is an electron,
its spin along each spatial axis $\vec{x}$ can take one of the two values $+\hbar/2$ and $-\hbar/2$
and, for each $\vec{x}$, `spin along $\vec{x}$ = $+\hbar/2$' and `spin along $\vec{x}$ = $-\hbar/2$'
represent two elements of the quantum logic.

In this paper, a quantum logic shall be an 
orthomodular partially ordered set $E$ 
with the partial ordering $\leq$,
the orthocomplementation $'$, 
the smallest element $0$ and the largest element $\mathbb{I}$
\cite{beltrametti1984logic, beran1985orthomodular, kalmbachorthomodular, ptak1991orthomodular}.
This means that the following conditions are satisfied by all $e,f \in E$:
\\[0,2 cm]
(A) $ e \leq f$ implies $f' \leq e'$.
\\[0,2 cm]
(B) $(e')' = e$.
\\[0,2 cm]
(C) $e \leq f'$ implies $e \vee f$, the supremum of $e$ and $f$, exists.
\\[0,2 cm]
(D) $e \vee e' = \mathbb{I}$.
\\[0,2 cm]
(E) $f \leq e$ implies $e = f \vee (e \wedge f')$. 
\hspace*{3,6 cm} (orthomodular law)
\\[0,2 cm]
Here, $e \wedge f$ denotes the infimum of $e$ and $f$, which exists if and only if $e' \vee f'$ exists.
Two elements $e,f \in E$ are called \textit{orthogonal} if $e \leq f'$ or, equivalently, $f \leq e'$.
An element $e \neq 0$ in $E$ is called an \textit{atom} if there is no element $f$ in $E$ 
with $f \leq e$ and $0 \neq f \neq e$. 

The interpretation of this mathematical terminology is as follows: 
two orthogonal elements represent mutually exclusive events, 
propositions or system properties; $e'$ is the negation of $e$, and
$e \vee f$ is the disjunction of the two exclusive elements $e$ and $f$.

It is not assumed that $E$ is a \textit{lattice} (in a lattice, 
there is a smallest upper bound $e \vee f$ and largest lower bound $e \wedge f$
for any two elements $e$ and $f$).
If $E$ were a lattice and satisfied the distributivity law
\begin{center}
$e \wedge (f \vee g) = (e \wedge f) \vee (e \wedge g)$ 
for all $e,f,g \in E$, 
\end{center}
$E$ would become a Boolean lattice or Boolean algebra. 
The orthomodular law is a weakening of the distributivity law. 

Classical probability theory uses Boolean algebras as mathematical structure 
for the random events, and it can be expected that those subsets of $E$,
which are Boolean algebras, behave classically. Therefore, a subset $E_0$ of $E$
is called \textit{compatible} if there is a Boolean algebra $B$ with $E_0 \subseteq B \subseteq E$.
Two elements $e$ and $f$ in $E$ are called compatible, if $\left\{e,f\right\}$ forms a compatible subset.
Note that the supremum $e \vee f$ and the infimum $e \wedge f$ exist 
for any compatible pair $e$ and $f$ in $E$
and that the distributivity law holds in any compatible subset of $E$.
Any subset with pairwise orthogonal elements is compatible \cite{brabec1979compatibility}.

Two subsets $E_1$ and $E_2$ of $E$ are called \textit{compatible with each other}
if the union of any compatible subset of $E_1$ with any compatible subset of $E_2$
is a compatible subset of $E$. Note that this does not imply that $E_1$ or $E_2$ themselves 
are compatible subsets.

A subset of an orthomodular \underline{lattice} is compatible if each pair of elements in this subset 
is compatible. However, the pairwise compatibility of the elements
of a subset of an orthomodular partially ordered set 
does not any more imply the compatibility of this subset 
\cite{brabec1979compatibility}.

A quantum logical structure, which is more general than the orthomodular partially ordered sets,
has been used in Refs. \cite{niestegge2001non, niestegge2008approach}. 
This more general structure is sufficient when only compatible pairs of elements in the 
quantum logic are considered. However, compatible subsets with more than two elements 
will play an important role in this paper. 

A quantum logic is a purely algebraic structure, unfurling its full potential 
only when its state space has some nice properties which shall be 
considered in the next section. 
\newpage
\noindent
\large
\textbf{3. Non-classical conditional probability}
\normalsize
\\[0,5 cm]
The states on the orthomodular partially ordered set $E$
are the analogue of the 
probability measures in classical probability theory, and 
conditional probabilities can be defined similar to 
their classical prototype. 

A \textit{state} $\rho$ allocates the probability $ \rho(f)$ 
with $0 \leq \rho(f) \leq 1$ to each element $f \in E$, 
is additive for orthogonal elements, and $\rho(\mathbb{I})=1$.
It then follows that $\rho(f) \leq \rho(e)$ for any two elements
$e,f \in E$ with $f \leq e$.

The \textit{conditional 
probability} of an element $f$ under another element $e$ is the 
updated probability for $f \in E$ after the outcome of
a first measurement has been $e \in E$; it is denoted 
by $ \rho(f | e) $. Mathematically, it is defined by the
conditions that the map $E \ni f \rightarrow \rho(f | e)$
is a state on $E$ and that it coincides with the classical
conditional probability for those $f$ which are compatible
with $e$; this means 
$$\rho(f | e) = \frac{\rho(e \wedge f)}{\rho(e)} ,$$
if $f$ is compatible with $e$.
It must be assumed that $\rho(e) \neq 0$.

However, among the orthomodular partially ordered sets, 
there are many where no states or no conditional 
probabilities exist, or where the conditional probabilities 
are ambiguous. It shall now be assumed 
for the remaining part of this paper that 
\begin{enumerate}
\item[(F)] there is a state $\rho$ on $E$ with $\rho(e)\neq 0$ for each $e \in E$ with $e \neq 0$,
\item[(G)] $E$ possesses unique conditional probabilities, and
\item[(H)] the state space of $E$ is strong; i.e., if 
$$\left\{ \rho \ |\  \rho \mbox{ is a state with } \rho(f) = 1 \right\} 
\subseteq \left\{ \rho \ |\  \rho \mbox{ is a state with } \rho(e) = 1 \right\}$$
holds for $e, f \in E$, then $f \leq e$. 
\end{enumerate}
If $\rho$ is a state with $\rho(e) = 1$ for some element $e \in E$, 
then $ \rho(f | e) = \rho(f)$ for all $f \in E$. This follows from (G).

For some pairs $e$ and $f$ in $E$, the conditional probability
does not depend on the underlying state; this means 
$\rho_1 (f|e) = \rho_2 (f|e)$ for all states $\rho_1$ and $\rho_2$
with $\rho_1 (e) \neq 0 \neq \rho_2 (e)$. This special conditional
probability is then denoted by $\mathbb{P} (f|e)$. The following 
two conditions are equivalent for any pair $e,f \in E$:
\\[0,3 cm]
\hspace*{0,2 cm}
(i) $\mathbb{P} (f|e)$ exists and $\mathbb{P} (f|e) = s$.
\\[0,3 cm]
\hspace*{0,2 cm}
(ii) $\rho(e) = 1$ implies $\rho(f) = s$ for the states $\rho$ on $E$. 
\\[0,3 cm]
Due to condition (H), $f \leq e$ holds for two elements $e$ and $f$ in $E$
if and only if $ \mathbb{P} (e|f) = 1$. Moreover, $e$ and $f$ are orthogonal
if and only if $ \mathbb{P} (e|f) = 0$.
 
$\mathbb{P} (f|e)$ exists for all $f \in E$ if and only if $e$ is an \textit{atom}
(minimal element in $E$), which results in the atomic state
$\mathbb{P}_e$ defined by $\mathbb{P}_e (f) := \mathbb{P} (f|e)$.
This is the unique state allocating the probability value $1$ to the atom $e$.

Quantum mechanics uses a special quantum logic; 
it consists of the self-adjoint projection operators on a 
Hilbert space $H$ and is an orthomodular lattice. Compatibility 
here means that the self-adjoint projection operators
commute. Conditions (F) and (H) are satisfied, and 
the unique conditional probabilities exist (G);
it has been shown in Ref. \cite{niestegge2001non} that,
with two self-adjoint projection operators $e$ and $f$ on $H$, the
conditional probability has the shape
$$\rho(f|e) = \frac{trace(aefe)}{trace(ae)} = \frac{trace(eaef)}{trace(ae)}$$
for a state $\rho$ defined by the statistical operator $a$ (i.e., $a$
is a self-adjoint operator on $H$ with non-negative spectrum and $trace(a)=1$). 
The above identity reveals that
conditionalization becomes identical with the state transition 
of the L\"uders - von Neumann measurement process. Therefore, 
the conditional probabilities can be 
regarded as a generalized mathematical model of projective quantum measurement.

$\mathbb{P}(f|e)$ exists with $\mathbb{P}(f|e) = s$ if and only if the operators
$e$ and $f$ on $H$ satisfy the algebraic identity $efe = se$.
This transition probability between the outcomes of two consecutive measurements
is independent of any underlying state and results from the algebraic structure
of the quantum logic.

The atoms are the self-adjoint projections on the one-dimensional subspaces 
of $H$; if $e$ is an atom and $| \xi \rangle $ a normalized vector in the 
corresponding one-dimensional subspace, then 
$\mathbb{P}(f|e) = \left\langle \xi| f \xi\right\rangle$.
The atomic states thus coincide with the quantum mechanical pure states or vector states.
Their general non-orthogonality is quite characteristic of quantum mechanics.  
If $f$ is an atom as well and $| \eta \rangle $ a normalized vector in the 
corresponding one-dimensional subspace, then 
$\mathbb{P}(f|e) = \left|\left\langle \eta| \xi\right\rangle\right|^{2}$.

Now it becomes evident why 
$\left|\left\langle \eta| \xi\right\rangle\right|^{2}$
represents a probability. In the usual quantum mechanical setting, 
this probabilistic interpretation (Born rule) is a later add-on
to the Hilbert space model, enforced by experimental evidence, but
rather unmotivated from the theoretical point of view.

In the case of the electron spin, with a fixed spatial axis $\vec{x}$, 
`spin along $\vec{x}$ = $+\hbar/2$' and `spin along $\vec{x}$ = $-\hbar/2$' 
represent two elements of the quantum logic which are orthogonal and compatible.
With two different axes $\vec{x}$ and $\vec{y}$, however, 
`spin along $\vec{x}$ = $+\hbar/2$' and `spin along $\vec{y}$ = $+\hbar/2$' 
represent two elements of the quantum logic which are not compatible.
The probability
\begin{center}
$\mathbb{P}($`spin along $\vec{x}$ = $+\hbar/2$'$|$`spin along $\vec{y}$ = $+\hbar/2$'$)$
\end{center}
exists; it depends on the angle $\theta$ between the two axes and equals $cos^{2} (\theta/2)$ ($ = 1/2$ for $\theta = \pi/2$). 
\newpage
\noindent
\large
\textbf{4. The generalized no-cloning theorem}
\normalsize
\\[0,5 cm]
In the remaining sections of this paper, a quantum logic 
shall always be an orthomodular partially ordered set 
satisfying (F), (G) and (H). This means that the existence 
of unique conditional probabilities is always assumed.

In the following, two lemmas from Ref.\cite{Nie2015PhysScrNoCloning} 
will be needed. The first one concerns the invariance of $\mathbb{P}(\ |\ )$
under morphisms and the second one the multiplicativity of $\mathbb{P}(\ |\ )$
under certain conditions. These lemmas shall be restated here without their
proofs which can be found in Ref.\cite{Nie2015PhysScrNoCloning}.

With two quantum logics $E$ and $F$, a \textit{morphism} is a map 
$T:E \rightarrow F$ satisfying the following three conditions:
(i) $Te_1 \leq Te_2$ for $e_1, e_2 \in E$ with $e_1 \leq e_2$, 
(ii) $T(e') = (Te)'$ for all $e \in E$, and 
(iii) $T \mathbb{I} = \mathbb{I}$.
\\[0,5 cm]
\textbf{Lemma 1:}
Suppose $E$ and $F$ are quantum logics
and $T:E \rightarrow F$ is a morphism.
If $\mathbb{P} \left( e_2| e_1 \right)$
exists for two elements $e_1$ and $e_2$ in $E$ with $T e_1 \neq 0$, then 
$\mathbb{P} \left( T e_2| T e_1 \right)$ exists and 
$\mathbb{P} \left( T e_2| T e_1 \right) = \mathbb{P} \left( e_2| e_1 \right).$
\\[0,5 cm]
Suppose that $E$ is a quantum logic and that two compatible copies of it are 
contained in the larger quantum logic $L$. This means that
there are two injective morphisms $\pi_a:E \rightarrow L$ and $\pi_b:E \rightarrow L$
and that the subsets $\pi_a(E)$ and $\pi_b(E)$ of $L$ 
are compatible with each other.
\\[0,5 cm]
\textbf{Lemma 2:}
If $\mathbb{P}(e_2|e_1)$ and $\mathbb{P}(f_2|f_1)$
both exist for $e_1, e_2, f_1, f_2 \in E$, then 
$$\mathbb{P}((\pi_a e_2) \wedge (\pi_b f_2)|(\pi_a e_1) \wedge (\pi_b f_1))$$
exists and
$$\mathbb{P}((\pi_a e_2) \wedge (\pi_b f_2)|(\pi_a e_1) \wedge (\pi_b f_1)) = \mathbb{P}(e_2|e_1) \mathbb{P}(f_2|f_1).$$

In the usual quantum mechanical setting, the cloning is performed 
by a unitary transformation on the Hilbert space tensor product. In this paper, it
shall be performed by an automorphism of $L$ containing the two compatible copies of $E$.
\\[0,5 cm]
\textbf{Theorem 1:}
Suppose that $e_1, e_2, f$ are elements of the quantum logic $E$,
that $\mathbb{P}(e_2|e_1)$ exists and
that $T:L \rightarrow L$ is an automorphism with 
$$T\left((\pi_a e_k) \wedge (\pi_b f)\right) = (\pi_a e_k) \wedge (\pi_b e_k)$$
for $k = 1,2$.
Then $\mathbb{P}(e_2|e_1) \in \left\{0,1\right\}$; 
this means that either $e_1$ and $e_2$ are orthogonal
or $e_1 \leq e_2$.
\newpage
\noindent
\textit{Proof}. By Lemma 1 and 2, on the one hand, 
\\[0,3 cm]
\hspace*{2,5 cm}
$\mathbb{P} \left( T\left((\pi_a e_2) \wedge (\pi_b f)\right) | \; T\left((\pi_a e_1) \wedge (\pi_b f)\right) \right)$
\\[0,3 cm]
\hspace*{4,0 cm}
$= \mathbb{P} \left( (\pi_a e_2) \wedge (\pi_b e_2) | (\pi_a e_1) \wedge (\pi_b e_1) \right) $
\\[0,3 cm]
\hspace*{4,0 cm}
$= \mathbb{P} (\pi_a e_2 | \pi_a e_1) \ \mathbb{P} (\pi_b e_2 | \pi_b e_1)$
\\[0,3 cm]
\hspace*{4,0 cm}
$= \left(\mathbb{P} (e_2|e_1)\right)^{2}$
\\[0,3 cm]
and, on the other hand,
\\[0,3 cm]
\hspace*{2,5 cm}
$\mathbb{P} \left( T\left((\pi_a e_2) \wedge (\pi_b f)\right) | \; T\left((\pi_a e_1) \wedge (\pi_b f)\right) \right)$
\\[0,3 cm]
\hspace*{4,0 cm}
$= \mathbb{P} \left( (\pi_a e_2) \wedge (\pi_b f) | \; (\pi_a e_1) \wedge (\pi_b f) \right)$
\\[0,3 cm]
\hspace*{4,0 cm}
$= \mathbb{P} (\pi_a e_2 | \pi_a e_1) \ \mathbb{P} (\pi_b f | \pi_b f)$
\\[0,3 cm]
\hspace*{4,0 cm}
$= \mathbb{P} (e_2|e_1).$
\\[0,3 cm]
Thus, $ \left(\mathbb{P} (e_2|e_1)\right)^{2} = \mathbb{P}(e_2|e_1) $ and therefore $\mathbb{P}(e_2|e_1) \in \left\{0,1\right\}$.
\newline

Theorem 1 means that the cloning or copying of the two unknown system properties $e_1, e_2$ 
becomes impossible, when $\mathbb{P}(e_2|e_1)$ exists and when this probability is different from 0 and 1.
If the system property is unknown in a larger set with more than two elements, the cloning is impossible
whenever this probability exists and is different from 0 and 1 for any pair of elements in this set.

$\mathbb{P}(e_2|e_1)$ always exists for atoms $e_1, e_2$ and, in this case, cloning is possible 
only if the atoms are orthogonal or identical. Atoms correspond to the atomic states 
(the generalization of the quantum mechanical pure states), and the cloning
of such atomic states was studied in Ref.\cite{Nie2015PhysScrNoCloning} which, however, required some 
additional assumptions that are not needed here. 

Theorem 1 is substantially more general 
than the result in Ref.\cite{Nie2015PhysScrNoCloning} and than the well-known quantum mechanical no-cloning theorem.
Even in the usual Hilbert space setting, Theorem 1 covers cases where $e_1$ and $e_2$ are not atomic 
and these cases are not included in other results. This becomes possible by considering the cloning of 
system properties instead of states and by using the specific probability $\mathbb{P}(\ |\ )$.

The original quantum mechanical no-cloning theorem \cite{wootters1982single, dieks1982communication} 
has been extended into different other directions:
to mixed states \cite{barnum1996noncommuting},
to C*-algebras \cite{clifton2003characterizing},
to finite-dimensional generic probabilistic models \cite{barnum2006cloning, barnum2007generalized}, and
to universal cloning \cite{:/content/aip/journal/jmp/50/10/10.1063/1.3245811}. Possible is only 
the approximate or imperfect cloning \cite{buvzek1996quantum, PhysRevA.57.2368, Kitajima2015}. 
However, none of these extensions covers the above result.
\newpage
\noindent
\large
\textbf{5. Quantum key distribution}
\normalsize
\\[0,5 cm]
\textbf{5.1 The first protocol}
\\[0,5 cm]
The quantum key distribution protocol 
invented by Charles H. Bennett and Gilles Brassard in 1984 \cite{BB84} 
shall be considered first, but instead of using the ordinary quantum 
mechanical Hilbert space formalism and pure states, it will be transferred 
to the more general setting described in the last three sections.  

Assume that two parties, usually named Alice and Bob, use a system, the properties of which 
form a quantum logic $E$ containing two elements $e$ and $f$ with 
$$\mathbb{P}(e|f) = \mathbb{P}(e|f') =\mathbb{P}(f|e) = \mathbb{P}(f|e') = 1/2.$$
Both agree in advance on using this system and these properties $e$ and $f$.
Alice encodes the bit 1 in either $e$ or $f$ and the bit 0 in either $e'$ or $f'$. 
She decides at random whether she uses $e, e'$ or $f, f'$ for the bit.
She then sends the system carrying the encoded bit to Bob. Theorem 1 ensures 
that an eavesdropper cannot make use of copying.

Bob does not know whether Alice used $e, e'$ or $f, f'$ and decides at random
whether he measures $e$ or $f$. In the case he measures $e$, the outcome $e$ 
means the bit 1 and the outcome $e'$ means the bit 0. In the case he measures $f$, the outcome $f$ 
means the bit 1 and the outcome $f'$ means the bit 0. When using the same property as Alice,
he gets the right bit with probability 1 
since $\mathbb{P}(e|e) =  \mathbb{P}(f|f) = 1$ 
and $\mathbb{P}(e|e') =  \mathbb{P}(f|f') = 0$; when using the other property he 
gets a useless random bit, since 
$\mathbb{P}(f|e) = \mathbb{P}(f|e') = \mathbb{P}(e|f) = \mathbb{P}(e|f')= 1/2$.

After Bob's measurement, he uses any classical and possibly public
communication channel to announce to Alice whether he has measured $e$ or $f$,
but he does not tell the result he obtained. Over the same communication channel, 
Alice then only announces whether or not
they both used the same one of the two system properties.
If they used the same one, they keep the bit; if not,they disregard it. 
Each of these two cases occurs with probability 1/2. 

After repeating the above 
procedure sufficiently many times with equal systems, 
Alice and Bob get a bit string which can they then use
as key for the encryption of messages which are then sent over the classical communication channel. 

Due to the no-cloning theorem in section 4, an eavesdropper cannot generate a copy of the signal 
which Alice sends to Bob without destroying the signal which Bob receives.
In this case, Bob would get the wrong bits (also when using the 
same property $e$ or $f$ as Alice) and the key would not work,  
since Alice and Bob use different bit strings. The eavesdropper can thus hinder 
the key distribution, but does not have access to the key without being detected. 

When the key 
works, Alice and Bob can be sure that nobody has a copy of the key.
They can check this by exchanging some unimportant test messages encrypted with the key 
over the classical channel, before starting to exchange secret information.
Another possibility is that they compare a small, but sufficiently large subset of their bit strings;
in the case of perfect coincidence, they can be sure that nobody has a copy of the key. If they 
cannot be sure, they discard the key and repeat the whole procedure to generate a new key.

Other and perhaps better procedures for this classical post-processing
have been proposed \cite{BB84, E91, gisin2002cryptoreview, QKD2004SARG, 2009cryptoreview}, 
but are not relevant for the scope of this paper.
\\[0,5 cm]
\textbf{5.2 A variant of the first protocol}
\\[0,5 cm]
Some time later, one of the co-inventors of the first protocol 
detected that it sufficient for Alice to use only the two non-compatible
quantum properties $e$ and $f$ instead of all the four
$e, e', f, f'$ and introduced the following variant \cite{bennett1992nonortho}.

Under the same assumptions as with the first protocol in subsection 5.1, 
Alice uses $e$ for encoding the bit 1 and $f$ for encoding the bit 0; 
she never uses $e'$ and $f'$. 

Bob again decides at random whether he measures $e$ or $f$. 
In the case he measures $e$, the outcome $e$ 
means the bit 1 and, in the case he measures $f$, the outcome $f$ 
means the bit 0. The other measurement outcomes $e'$ and $f'$ are useless and
result from two different reasons; 
either Bob's random selection among $e$ and $f$ does not coincide 
with Alice's coding or 
the transmission was not correct because of the presence of an eavesdropper 
or some transmission problems.

The procedure is then continued in the same way as described in the last subsection.
\\[0,5 cm]
\textbf{5.3 The second protocol}
\\[0,5 cm]
A further quantum key distribution protocol invented by
Artur K. Ekert in 1991 \cite{E91} shall now be considered.
The major difference to the first one is that it uses 
two entangled subsystems for the transmission of each bit. Alice and 
Bob can exchange these subsystems in advance, before Alice encodes the bit. 
After this, no system is transmitted from Alice to Bob.

Consider a quantum logic $E$ with five elements $e_a, f_a, e_b, f_b, d$ 
satisfying the following three conditions:
\begin{enumerate}
\item[(i)]
The subsets $\left\{e_a, e_a', f_a, f_a'\right\}$ and $\left\{e_b, e_b',f_b, f_b'\right\}$ are compatible with each other.
\item[(ii)]
$\mathbb{P}(e_k|f_k) = \mathbb{P}(e_k|f_k') =\mathbb{P}(f_k|e_k) = \mathbb{P}(f_k|e_k') = 1/2$ for $k = a,b$.
\item[(iii)]
$\mathbb{P}(e_k|d) = \mathbb{P}(f_k|d) = 1/2$ for $k = a,b$ and 
$\mathbb{P}(e_a \wedge e_b|d) = \mathbb{P}(f_a \wedge f_b|d) = 1/2$,
$\mathbb{P}(e_a \wedge f_b|d) = \mathbb{P}(f_a \wedge e_b|d) = 1/4$.
\end{enumerate}
Note that (iii) implies some further identities. From  
$\mathbb{P}(e_a|d)= \mathbb{P}(e_a \wedge e_b|d) + \mathbb{P}(e_a \wedge e_b'|d)$,
it follows that $\mathbb{P}(e_a \wedge e_b'|d) = 0$. The identities
$\mathbb{P}(e_a' \wedge e_b|d) = 0$,  $\mathbb{P}(f_a \wedge f_b'|d) = 0$ and  $\mathbb{P}(f_a' \wedge f_b|d) = 0$
follow in the same way. The orthogonal decomposition 
$ \mathbb{I} = (e_a \wedge e_b) \vee (e_a \wedge e_b') \vee (e_a' \wedge e_b) \vee (e_a' \wedge e_b')$
then implies 
$$\mathbb{P}(e_a' \wedge e_b'|d) = 1/2$$ 
and
$$\mathbb{P}(f_a' \wedge f_b'|d) = 1/2$$
follows in the same way. Moreover, 
$\mathbb{P}(e_a|d) = \mathbb{P}(e_a \wedge f_b|d) + \mathbb{P}(e_a \wedge f_b'|d)$
implies 
$$\mathbb{P}(e_a \wedge f_b'|d) = 1/4.$$
In the same way, it is concluded that
$$1/4 = \mathbb{P}(e_a' \wedge f_b|d) = \mathbb{P}(e_a' \wedge f_b'|d) =  
\mathbb{P}(f_a \wedge e_b'|d) = \mathbb{P}(f_a' \wedge e_b|d) = 
\mathbb{P}(f_a' \wedge e_b'|d).$$

In the beginning of the key distribution process, 
the system is prepared in such a way that property $d$ holds. The subsets
$\left\{e_a, e_a', f_a, f_a'\right\}$ and $\left\{e_b, e_b',f_b, f_b'\right\}$ 
represent two subsystems; the first one is Alice's
and the second one is Bob's.

The further proceeding is similar to the first quantum key distribution protocol.
The only difference is that Alice and Bob use their different subsystems instead
of transmitting the system from Alice to Bob.

However, it is more difficult to see that Bob gets the same bit as Alice when 
she uses $e_a, e_a'$ and he $e_b, e_b'$ or when
she uses $f_a, f_a'$ and he $f_b, e_b'$.
Assume that Alice encodes the bit 1 using $e_a$ and that Bob measures $e_b$. 
Since $e_a$ and $e_b$ are compatible, the conditional probability that 
Bob gets $e_b$ after Alice got $e_a$ is
$$\frac{\mathbb{P}(e_a \wedge e_b|d)}{\mathbb{P}(e_a|d)}$$
and equals 1 because of condition (iii).
The same holds for the pairs $e_a'$ and $e_b'$, $f_a$ and $f_b$, $f_a'$ and $f_b'$,
using condition (iii) and the identities following from it.
This means that Bob gets the right bit, when his random choice between $e$ and $f$
coincides with Alice's choice. 

When their choices don't match, he gets a useless
random bit, 0 or 1, each with equal probability 1/2.  
Assume that Alice encodes the bit 1 using $e_a$ 
and that Bob measures $f_b$. Then 
$$\frac{\mathbb{P}(e_a \wedge f_b|d)}{\mathbb{P}(e_a|d)}$$
equals $1/2$ because of condition (iii).
The same holds for the pairs $e_a$ and $f_b'$, $e_a'$ and $f_b$, $e_a'$ and $f_b'$,
$f_a$ and $e_b$, $f_a$ and $e_b'$, $f_a'$ and $e_b$, $f_a'$ and $e_b'$,
using condition (iii) and the identities following from it.

The key distribution procedure is continued in the same way as described in subsection 5.1
for the first protocol.
\\[0,5 cm]
\textbf{5.4 The usual quantum mechanical Hilbert space model}
\\[0,5 cm]
The quantum key distribution protocols shall now be studied in
the special quantum logic consisting of the self-adjoint
projection operators on a Hilbert space $H$. A reader familiar with the
protocols will now recognize them if not yet in subsections 5.1, 5.2 and 5.3.

Consider a 2-dimensional Hilbert space $H$ with an orthonormal basis
$|\xi_1\rangle$ and $|\xi_2\rangle$. A second orthonormal basis 
is then given by 
\begin{center}
$|\eta_1\rangle := \frac{1}{\sqrt{2}} \left( |\xi_1\rangle + |\xi_2\rangle \right)$ 
and $|\eta_2\rangle := \frac{1}{\sqrt{2}} \left( |\xi_1\rangle - |\xi_2\rangle \right)$. 
\end{center}
Now define the following self-adjoint projection operators: 
\begin{center}
$ e := | \eta_1 \rangle \langle \eta_1 |$ and $ f := | \xi_1 \rangle \langle \xi_1 |$.
\end{center}
They satisfy the assumptions needed for the first protocol and its variant: 
\begin{center}
$\mathbb{P}(e|f) = \mathbb{P}(e|f') =\mathbb{P}(f|e) = \mathbb{P}(f|e') = 1/2$.
\end{center}
Furthermore, it here becomes evident that the situation, where 
$\mathbb{P}(\ |\ )$ exists with $0 \neq \mathbb{P}(\ |\ ) \neq 1$,
is an extension of the superposition principle of Hilbert space quantum mechanics
to a much more general setting.

For the second protocol, consider the tensor product $H \otimes H$, its element 
\begin{center}
$ | \psi \rangle := \frac{1}{\sqrt{2}} \left( | \xi_1 , \xi_1 \rangle + | \xi_2 , \xi_2 \rangle\right) \in H \otimes H $
\end{center}
and define  
\begin{center}
$ d := | \psi \rangle \langle \psi |$, 
\end{center}
\begin{center}
$e_a := e \otimes \mathbb{I}$, $f_a := f \otimes \mathbb{I}$, 
\end{center}
\begin{center}
$e_b := \mathbb{I} \otimes e$, $f_b := \mathbb{I} \otimes f$,
\end{center}
where $\mathbb{I}$ is the identity operator on $H$. 
They satisfy the conditions (i), (ii) and (iii), needed for the second protocol. 

The element $| \psi \rangle $ 
in the tensor product of the two Hilbert spaces
represents a typical entangled state.
A different way of thinking,
suggested by the approach presented in this paper,
is to attribute the entanglement not to states,
but to system properties and, in this special case, to the element 
$d = | \psi \rangle \langle \psi |$ of the quantum logic; it represents
a property of the total system which cannot be described by the properties of the 
two individual subsystems.

In the physics literature, the elements 
of the Hilbert space $H$ are often called \textit{wavefunctions}.
They don't have any equivalent in the general and abstract setting
provided by the quantum logics possessing a conditional probability calculus.
Even in usual quantum mechanical setting,
the key distribution protocols 
(and the no-cloning theorem)
can do without the wavefunctions;
instead, the elements of the Hilbert space quantum logic (the self-adjoint projection operators)
together with the special state-independent conditional probability $\mathbb{P}(\ |\ )$
can be used. This probability, representing the transition probability between
the measurement outcomes, then replaces the 
measurement induced so-called collapse of the wavefunction
and becomes the quantum origin of the no-cloning theorem as well as of 
the superior security of the
key distribution protocols.

\newpage
\noindent
\large
\textbf{6. Conclusion}
\normalsize
\\[0,5 cm]
In this paper, it has been shown that
the quantum no-cloning theorem and quantum key distribution
allow a much more general and abstract access
than commonly thought. Instead of 
the usual quantum mechanical Hilbert space model,
a general non-classical extension of conditional 
probability has been used, which
includes the usual model as a special case. 

Equally important may be that,  
even in usual quantum mechanics, more cases are covered, 
since any elements in the quantum logic and not only the atoms
(which correspond to the pure states) can be used
for quantum key distribution. 
This then includes the quantum logics
formed by the self-adjoint idempotent elements of the type II or type III
von Neumann factors which occur naturally in relativistic quantum field theory
and in quantum statistical
mechanics of infinite systems \cite{yngvason2005role_typeIII,sakai2012book}.
These quantum logics do not contain any atom and do not possess
any normal pure state, but there are elements $e$ and $f$ with $0 < \mathbb{P}(f|e) < 1$. 

This special type of conditional probability $\mathbb{P}(f|e)$
could be identified as the quantum origin of the no-cloning theorem and 
the superior security of the key distribution protocols considered.
The probability $\mathbb{P}(f|e)$ is a transition probability 
between the two events, propositions or system properties $e$ and $f$. 
It depends only on $e$ and $f$ and does not depend on any underlying quantum state.
It results solely from the algebraic structure of the quantum logic.

This means that the usual assumption that every quantum system
(e.g., a microphysical particle or the whole universe)
is in a quantum state is not necessary any more
to understand the probabilities observed in the quantum physical realm.
No wavefunction and no collapse of it is needed.
Some interpretational problems with quantum mechanics,
resulting from this usual assumption, might thus be removed 
or shifted to another kind of questions. 
So far, the special type of conditional probability $\mathbb{P}(\ |\ )$
has not attracted much attention, although it may have more
profound implications for the foundations and interpretation of quantum theory,
quantum information theory and the philosophical
question, what actually constitutes physical reality.

Only the two oldest and best-known quantum key distribution protocols 
(including a variant) have been considered in the present paper,
although there are more \cite{gisin2002cryptoreview, 2009cryptoreview}.
Moreover, quantum key distribution is only one of several tasks in quantum information theory. 
Others are entanglement-assisted quantum teleportation \cite{bennett1993teleporting}, 
quantum computing and specific quantum algorithms
like Grover's search algorithm \cite{grover1996, grover1997} 
or Shor's algorithm for integer factorization \cite{shor1994}.
The question, whether these tasks allow the same general and abstract access 
as the quantum key distribution protocols considered here,
will be tackled in a forthcoming paper. 
The essential feature 
needed for all of them is quantum entanglement and, that this feature 
can be transferred to the general and abstract setting, has already been demonstrated successfully 
in the treatment of the second quantum distribution protocol.
A new way of thinking required here is to attribute the entanglement not to states,
but to system properties.

\bibliographystyle{abbrv}
\bibliography{Literatur}

\begin{thebibliography}{10}

\bibitem{barnum2006cloning}
H.~Barnum, J.~Barrett, M.~Leifer, and A.~Wilce.
\newblock Cloning and broadcasting in generic probabilistic theories.
\newblock {\em arXiv:quant-ph/0611295}, 2006.

\bibitem{barnum2007generalized}
H.~Barnum, J.~Barrett, M.~Leifer, and A.~Wilce.
\newblock Generalized no-broadcasting theorem.
\newblock {\em Physical Review Letters}, 99(24):240501, 2007.

\bibitem{barnum1996noncommuting}
H.~Barnum, C.~M. Caves, C.~A. Fuchs, R.~Jozsa, and B.~Schumacher.
\newblock Noncommuting mixed states cannot be broadcast.
\newblock {\em Physical Review Letters}, 76(15):2818, 1996.

\bibitem{barrett2007information}
J.~Barrett.
\newblock Information processing in generalized probabilistic theories.
\newblock {\em Physical Review A}, 75(3):032304, 2007.

\bibitem{beltrametti1984logic}
E.~G. Beltrametti, G.~Cassinelli, and G.-C. Rota.
\newblock {\em The logic of quantum mechanics}.
\newblock Cambridge University Press, 1984.

\bibitem{bennett1992nonortho}
C.~H. Bennett.
\newblock Quantum cryptography using any two nonorthogonal states.
\newblock {\em Physical Review Letters}, 68(21):3121, 1992.

\bibitem{BB84}
C.~H. Bennett and G.~Brassard.
\newblock Quantum cryptography: {P}ublic key distribution and coin tossing.
\newblock In {\em Proceedings of IEEE International Conference on Computers,
  Systems and Signal Processing (Bangalore, India, Dec. 1984)}, volume 175,
  page~8, 1984.

\bibitem{bennett1993teleporting}
C.~H. Bennett, G.~Brassard, C.~Cr{\'e}peau, R.~Jozsa, A.~Peres, and W.~K.
  Wootters.
\newblock Teleporting an unknown quantum state via dual classical and
  {Einstein-Podolsky-Rosen} channels.
\newblock {\em Physical Review Letters}, 70(13):1895, 1993.

\bibitem{beran1985orthomodular}
L.~Beran.
\newblock {\em Orthomodular lattices}.
\newblock Springer, 1985.

\bibitem{birkhoff-vN36}
G.~Birkhoff and J.~von Neumann.
\newblock The logic of quantum mechanics.
\newblock {\em Annals of Mathematics}, 37:823--843, 1936.

\bibitem{brabec1979compatibility}
J.~Brabec.
\newblock Compatibility in orthomodular posets.
\newblock {\em {\v{C}}asopis pro p{\v{e}}stov{\'a}n{\'\i} matematiky},
  104(2):149--153, 1979.

\bibitem{PhysRevA.57.2368}
D.~Bru\ss{}, D.~P. DiVincenzo, A.~Ekert, C.~A. Fuchs, C.~Macchiavello, and
  J.~A. Smolin.
\newblock Optimal universal and state-dependent quantum cloning.
\newblock {\em Phys. Rev. A}, 57:2368--2378, Apr 1998.

\bibitem{bub1977neumann}
J.~Bub.
\newblock Von {Neumann's} projection postulate as a probability
  conditionalization rule in quantum mechanics.
\newblock {\em Journal of Philosophical Logic}, 6(1):381--390, 1977.

\bibitem{buvzek1996quantum}
V.~Bu{\v{z}}ek and M.~Hillery.
\newblock Quantum copying: {Beyond} the no-cloning theorem.
\newblock {\em Physical Review A}, 54(3):1844, 1996.

\bibitem{clifton2003characterizing}
R.~Clifton, J.~Bub, and H.~Halvorson.
\newblock Characterizing quantum theory in terms of information-theoretic
  constraints.
\newblock {\em Foundations of Physics}, 33(11):1561--1591, 2003.

\bibitem{dieks1982communication}
D.~Dieks.
\newblock Communication by {EPR} devices.
\newblock {\em Physics Letters A}, 92(6):271--272, 1982.

\bibitem{edwards1990conditional}
C.~M. Edwards and G.~T. R{\"u}ttimann.
\newblock On conditional probability in {GL} spaces.
\newblock {\em Foundations of Physics}, 20(7):859--872, 1990.

\bibitem{E91}
A.~K. Ekert.
\newblock Quantum cryptography based on {B}ell's theorem.
\newblock {\em Phys. Rev. Lett.}, 67:661--663, Aug 1991.

\bibitem{friedman1978quantum}
M.~Friedman and H.~Putnam.
\newblock Quantum logic, conditional probability, and interference.
\newblock {\em Dialectica}, 32(3-4):305--315, 1978.

\bibitem{gisin2002cryptoreview}
N.~Gisin, G.~Ribordy, W.~Tittel, and H.~Zbinden.
\newblock Quantum cryptography.
\newblock {\em Reviews of Modern Physics}, 74(1):145, 2002.

\bibitem{grover1996}
L.~K. Grover.
\newblock A fast quantum mechanical algorithm for database search.
\newblock In {\em Proceedings of the twenty-eighth annual ACM symposium on
  Theory of computing}, pages 212--219. ACM, 1996.

\bibitem{grover1997}
L.~K. Grover.
\newblock Quantum mechanics helps in searching for a needle in a haystack.
\newblock {\em Physical Review Letters}, 79(2):325, 1997.

\bibitem{gunson1967algebraic}
J.~Gunson.
\newblock On the algebraic structure of quantum mechanics.
\newblock {\em Communications in Mathematical Physics}, 6(4):262--285, 1967.

\bibitem{guz1981conditional}
W.~Guz.
\newblock Conditional probability and the axiomatic structure of quantum
  mechanics.
\newblock {\em Fortschritte der Physik}, 29(8):345--379, 1981.

\bibitem{kalmbachorthomodular}
G.~Kalmbach.
\newblock {\em Orthomodular lattices}.
\newblock Academic Press, London, 1983.

\bibitem{Keller1980}
H.~A. Keller.
\newblock Ein nicht-klassischer {Hilbertscher Raum}.
\newblock {\em Mathematische Zeitschrift}, 172(1):41--49, 1980.

\bibitem{Kitajima2015}
Y.~Kitajima.
\newblock Imperfect cloning operations in algebraic quantum theory.
\newblock {\em Foundations of Physics}, 45(1):62--74, 2015.

\bibitem{:/content/aip/journal/jmp/50/10/10.1063/1.3245811}
T.~Miyadera and H.~Imai.
\newblock No-cloning theorem on quantum logics.
\newblock {\em Journal of Mathematical Physics}, 50(10):--, 2009.

\bibitem{Nie1998HPA01}
G.~Niestegge.
\newblock Statistische und deterministische vorhersagbarkeit bei der
  quantenphysikalischen messung.
\newblock {\em Helvetica Physica Acta}, 71(2):163--183, 1998.

\bibitem{niestegge2001non}
G.~Niestegge.
\newblock Non-{Boolean} probabilities and quantum measurement.
\newblock {\em Journal of Physics A: Mathematical and General}, 34(30):6031,
  2001.

\bibitem{niestegge2008approach}
G.~Niestegge.
\newblock An approach to quantum mechanics via conditional probabilities.
\newblock {\em Foundations of Physics}, 38(3):241--256, 2008.

\bibitem{Nie2015PhysScrNoCloning}
G.~Niestegge.
\newblock Non-classical conditional probability and the quantum no-cloning
  theorem.
\newblock {\em Physica Scripta}, 90(9):095101, 2015.

\bibitem{piron1964axiomatique}
C.~Piron.
\newblock Axiomatique quantique.
\newblock {\em Helvetica Physica Acta}, 37(4-5):439--468, 1964.

\bibitem{ptak1991orthomodular}
P.~Pt{\'a}k and S.~Pulmannov{\'a}.
\newblock {\em Orthomodular structures as quantum logics}.
\newblock Kluwer, Dordrecht, 1991.

\bibitem{sakai2012book}
S.~Sakai.
\newblock {\em C*-algebras and W*-algebras}.
\newblock Springer Science \& Business Media, 2012.

\bibitem{QKD2004SARG}
V.~Scarani, A.~Acin, G.~Ribordy, and N.~Gisin.
\newblock Quantum cryptography protocols robust against photon number splitting
  attacks for weak laser pulse implementations.
\newblock {\em Physical Review Letters}, 92(5):057901, 2004.

\bibitem{2009cryptoreview}
V.~Scarani, H.~Bechmann-Pasquinucci, N.~J. Cerf, M.~Du{\v{s}}ek,
  N.~L{\"u}tkenhaus, and M.~Peev.
\newblock The security of practical quantum key distribution.
\newblock {\em Reviews of Modern Physics}, 81(3):1301, 2009.

\bibitem{shor1994}
P.~W. Shor.
\newblock Algorithms for quantum computation: Discrete logarithms and
  factoring.
\newblock In {\em Foundations of Computer Science, 1994 Proceedings., 35th
  Annual Symposium on}, pages 124--134. IEEE, 1994.

\bibitem{soler1995characterization}
M.~P. Soler.
\newblock Characterization of {Hilbert} spaces by orthomodular spaces.
\newblock {\em Communications in Algebra}, 23(1):219--243, 1995.

\bibitem{varadarajan1968geometry}
V.~S. Varadarajan.
\newblock {\em Geometry of Quantum Theory, Vol. 1}.
\newblock Van Nostrand - Reinhold, New York, 1968.

\bibitem{varadarajan1970geometry}
V.~S. Varadarajan.
\newblock {\em Geometry of Quantum Theory, Vol. 2}.
\newblock Van Nostrand - Reinhold, New York, 1970.

\bibitem{wiesner1983}
S.~Wiesner.
\newblock Conjugate coding.
\newblock {\em ACM Sigact News}, 15(1):78--88, 1983.

\bibitem{wootters1982single}
W.~K. Wootters and W.~H. Zurek.
\newblock A single quantum cannot be cloned.
\newblock {\em Nature}, 299(5886):802--803, 1982.

\bibitem{yngvason2005role_typeIII}
J.~Yngvason.
\newblock The role of type {III} factors in quantum field theory.
\newblock {\em Reports on Mathematical Physics}, 55(1):135--147, 2005.

\bibitem{yuen1986amplification}
H.~P. Yuen.
\newblock Amplification of quantum states and noiseless photon amplifiers.
\newblock {\em Physics Letters A}, 113(8):405--407, 1986.

\end{thebibliography}

\end{document}